\documentclass[%
 aip,
 amsmath,amssymb,
 reprint,%
]{revtex4-1}

\usepackage{graphicx}
\usepackage{dcolumn}
\usepackage{bm}

\usepackage[utf8]{inputenc}
\usepackage[T1]{fontenc}
\usepackage{mathptmx}
\usepackage{etoolbox}

\newcommand{\zbar}{\overline{z}}
\newcommand{\abar}{\overline{\alpha}}

\makeatletter
\def\@email#1#2{%
 \endgroup
 \patchcmd{\titleblock@produce}
  {\frontmatter@RRAPformat}
  {\frontmatter@RRAPformat{\produce@RRAP{*#1\href{mailto:#2}{#2}}}\frontmatter@RRAPformat}
  {}{}
}%
\makeatother
\begin{document}

\preprint{AIP/123-QED}

\title[Edge modes of the Helmholtz-Onsager gas in a multiply-connected domain]{Edge modes of the Helmholtz-Onsager gas in a multiply-connected domain}
\author{Richard McQueen II}
  \thanks{Corresponding author}
  \email{mcquer@edu.}
\author{Chjan C. Lim}%
\affiliation{ 
Rensselaer Polytechnic Institute, Mathematical Sciences Dept., Troy, NY 12180, United States
}%

\date{\today}

\begin{abstract}
The vortex gas is an approximation used to study 2D flow using statistical mechanics methodologies. We investigate low positive Onsager temperature states for the vortex gas on an annular domain. Using mean field theory, microcanonical sampling of the point gas model, and canonical sampling of a lattice model, we find evidence for edge modes at low energy states.
\end{abstract}

\maketitle

\section{\label{sec:intro} Introduction}

The vortex gas is a class of solutions to the 2D Euler equations, in which vorticity is approximated by a collection of point particles whose positions form a Hamiltonian system. Vortices are relevant to many models of plasma, in both astrophysical phenomena, as well as confined settings such as tokamaks, lasers, and etching devices. They are observed on multiple scales; for instance they can exist in magnetized plasma on the length scale of the electron skin depth if this value is small enough, or on macroscopic scales\cite{skin-depth, netherlands-two-fluid, coherent-high-temp}. Vortices can be seen in drift waves\cite{christiansen, ussr-nonlinear-rossby, nycander-steady}, the Hasegawa-Mima equation\cite{stationary-hm, france-ukraine-2D-plasma}, 2D magnetohydrodynamics\cite{mhd}, and acoustic waves\cite{gas-discharge}. \\

Using the vortex gas model allows a statistical mechanics framework to be applied\cite{neutral-gas, thermometry, lim-assad-planar, extremal-rotating}. In the particular case of a bounded domain the state space for point vortices is also bounded, which means that the energy distribution across all configurations is unimodal and therefore one can consider either positive or negative statistical mechanical temperature\cite{u-stats, 2d-stat-mech}. Much work\cite{lim-ding, lim-neg-temp-sphere, neg-entropy, v-bouchet}, has already been done in the high kinetic energy/negative temperature regime. \\

Less work, however, has been done in the low kinetic energy/positive temperature regime. We seek here to expand on the work of Patil and Dunkel\cite{patil-dunkel} in 2021, which found evidence for edge modes in the vortex gas at low temperatures in both convex and concave domains. We expand this result to a multiply-connected domain, implying that this phenomenon would hold in a wide variety of spaces. This is achieved by three approaches. \\

The Sinh-Poisson mean field equation\cite{gurarie-chow} that arises from the point vortex model can be solved on subsets of the plane, including the annulus. To support the analytical findings we sample low energy distributions of point vortices and using the Green's function for the domain simulate their time evolution, before combining all of this data into a density. In both cases we find evidence of edge states at low temperatures. \\

In addition to point vortex simulations, we investigate the use of lattice models\cite{mc-phase-transitions, lim-nebus, long-range, newton-vortex} to analyze vorticity on the annular domain. Due to potential inequivalence between ensembles for non-additive systems such as the vortex gas\cite{bouchet-ens-diff}, this third approach makes use of the canonical ensemble. The lattice models show advantages in terms of computing time, while also having drawbacks with precision and range of applicability. The findings of these models regarding edge states largely corroborate the continuum case. \\

\section{Background}

We start with the Onsager vortex gas model\cite{onsager-primary}. Vorticities move according to a Hamiltonian system
\begin{subequations}
\begin{equation}\label{kirchhoff}
    \mathcal{H} = -\sum_{a,b} \lambda_{a} \lambda_{b} G(\mathbf{r}_{a}, \mathbf{r}_{b}) - \sum_{a} \lambda_{a}^{2} g(\mathbf{r}_{a}).
\end{equation}
\begin{equation}
    \frac{ \partial x_{a}}{ \partial t} = \frac{ \partial \mathcal{H}}{ \partial y_{a}}; \;\;\;\;\;\;\;\; \frac{ \partial y_{a}}{ \partial t} = -\frac{ \partial \mathcal{H}}{ \partial x_{a}}
\end{equation}
\end{subequations}
The differential system can also be written in a condensed form if we identify $\mathbf{r}_{a} = (x_{a}, y_{a})$ with a complex variable $z_{a} = x_{a} + iy_{a}$
\begin{equation}
    \frac{ \partial z_{a}}{ \partial t} = \frac{\partial \mathcal{H}}{ \partial z_{a}^{*}}
\end{equation}

The overall Hamiltonian $\mathcal{H}$ corresponds to the kinetic energy of the underlying fluid. The energy of two vortices of unit strength at $\mathbf{r}_{a}$ and $\mathbf{r}_{b}$ is given by the domain's Green's function $G(\mathbf{r}_{a}; \mathbf{r}_{b}) = G(\mathbf{r}_{b}; \mathbf{r}_{a})$. The single-index summation that appears in \eqref{kirchhoff} consists of self-interaction terms. The function $g$, sometimes referred to as the Robin function, is defined as the ``difference'' between two infinite quantities and reflects the geometry of the domain,\cite{robin-book, vortex-robin}
\begin{equation}\label{robin}
    g(\mathbf{r}) = \lim_{ \mathbf{z} \rightarrow \mathbf{r}} \big{(} G(\mathbf{z}, \mathbf{r}) - \frac{1}{2\pi} \log \vert \mathbf{z} - \mathbf{r} \vert \big{)}.
\end{equation}
The logarithm is the Green's function for the Laplace operator on the plane; subtracting it out can be thought of as removing the infinite self energy of the vortex. \\

We are interested in the large $N$ and long-time behavior of this model, on an annular domain with one vortex species.

\section{Annulus}

Consider the annulus centered at the origin with an outer radius of $1$ and an inner radius of $q$. In this paper, $q$ will always be equal to $0.3$. The Green's function for this domain is given by \cite{crowdy-marshall}
\begin{eqnarray}\label{annulus-greens}
    G(z, \alpha) &=& \frac{1}{2\pi} \log \Big{|} \frac{ (z - \alpha) \prod_{k=1}^{ \infty} (1 - q^{2k} \frac{z}{ \alpha})(1 - q^{2k} \frac{ \alpha}{z}) }{ (1 - \alpha^{*}z) \prod_{k=1}^{ \infty} (1 - q^{2k} \alpha^{*}z)(1 - q^{2k} (\alpha^{*}z)^{-1}) } \Big{|} \nonumber \\
    &&+ \frac{1}{2\pi} \frac{ \log \vert \alpha \vert \cdot \log \vert z \vert }{ \log(q) }.
\end{eqnarray}
And Eq.~(\ref{robin}) becomes
\begin{eqnarray}
    g(z) &=& \frac{1}{2\pi} \log \Big{|} \frac{ \prod_{k=1}^{ \infty} (1 - q^{2k})^{2} }{ (1 - \vert z \vert^{2}) \prod_{k=1}^{ \infty} (1 - q^{2k} \vert z \vert^{2})(1 - q^{2k} \vert z \vert^{-2}) } \Big{|} \nonumber \\
    &&+ \frac{1}{2\pi} \frac{ (\log \vert z \vert)^{2} }{ \log(q) }.
\end{eqnarray}
We can investigate the derivatives of these functions analytically:
\begin{eqnarray}
    4 \pi \frac{ \partial G}{ \partial \zbar} = \frac{1}{\zbar - \abar} + \sum_{k=1}^{ \infty} \frac{1}{ \zbar - q^{-2k} \abar } + \sum_{k=1}^{ \infty} \frac{1}{ \zbar^{2}} \frac{ q^{2k} \abar }{ 1 - q^{2k} \frac{ \abar}{\zbar} } \nonumber \\
    - \frac{1}{\zbar - \alpha^{-1}} - \sum_{k=0}^{ \infty} \frac{1}{\zbar - q^{-2k} \alpha^{-1}} - \sum_{k=1}^{ \infty} \frac{1}{ \zbar^{2}} \frac{ q^{2k} \abar^{-1} }{ 1 - q^{2k} \frac{ \abar^{-1}}{ \zbar} } + \frac{ \log \vert \alpha \vert}{ \log(q)} \frac{1}{ \zbar} \nonumber \\
    = \frac{1}{\zbar - \abar} + \sum_{k=1}^{ \infty} \frac{1}{ \zbar - q^{-2k} \abar } + \frac{ \abar}{ \zbar} \sum_{k=1}^{ \infty} \frac{ q^{2k}}{ \zbar - q^{2k} \abar } \nonumber \\
    - \frac{1}{\zbar - \alpha^{-1}} - \sum_{k=1}^{ \infty} \frac{1}{\zbar - q^{-2k} \alpha^{-1}} - \frac{ \abar^{-1}}{ \zbar} \sum_{k=1}^{ \infty} \frac{ q^{2k} }{ \zbar - q^{2k} \abar^{-1} } + \frac{ \log \vert \alpha \vert}{ \log(q)} \frac{1}{ \zbar}
\end{eqnarray}
And
\begin{eqnarray}
    2\pi \frac{ \partial g}{ \partial \zbar} &=& \frac{2}{\log(q)} \frac{ \log \vert z \vert}{ \zbar} + \frac{z}{1 - \vert z \vert^{2}} \nonumber \\
    &&+ \sum_{k=1}^{ \infty} \frac{ q^{2k}z}{ 1-q^{2k}\vert z \vert^{2}} - \sum_{k=1}^{ \infty} \frac{1}{z \zbar^{2}} \frac{ q^{2k}}{1 - q^{2k} \frac{1}{z \zbar}} \nonumber \\
    &=& \frac{2}{\log(q)} \frac{ \log \vert z \vert}{ \zbar} - \frac{1}{\zbar - z^{-1}} \nonumber \\
    &&- \sum_{k=1}^{ \infty} \frac{1}{ \zbar - q^{-2k}z^{-1} } - \frac{1}{\vert z \vert^{2}} \sum_{k=1}^{ \infty} \frac{ q^{2k}}{\zbar - q^{2k} z^{-1}}
\end{eqnarray}
In the simply connected disc, an interpretation of the Hamiltonian is as having ``image vortices''\cite{patil-dunkel} of opposite sign located at $\zbar^{-1}$, which gives rise to boundary effects at low temperatures. We see something similar here. The expression is suggestive of infinitely many image points (the ones near the origin being of decreasing intensity), but of primary importance are the three terms that cannot be bounded: $(\zbar - \abar)^{-1}$, $-(\zbar - \alpha^{-1})^{-1}$, and $-q^{2} (\zbar - q^{2} \alpha^{-1})^{-1}$. The first corresponds to the normal planar interaction between vortices, the second corresponds to image vortices which can potentially be located very close to the outer boundary of the domain, and the third gives an image vortex that can be located very close to the inner circle with radius $q$. \\

A perspective from which one can view the system is that the self-interaction terms look and act like a magnetic field term in an Ising model\cite{berlin-kac, pathria}. When the system's temperature is high, the system is dominated by proximity between vortices being high energy. At lower temperatures there is a phase transition and the vortices settle into the low energy regions imposed by the domain.

\section{Mean Field Density}

The existence and validity of mean field solutions for point vortex systems were found by \cite{kiessling} and \cite{caglioti}. \\

The Boltzmann entropy\cite{gibbs-v-boltz} is defined as
\begin{equation}
    S = -\int_{ \Omega} \rho \log \rho \, d\mathbf{r},
\end{equation}
where $\rho$ is a vortex density. \\

To find the mean field density, we maximize $S$ subject to energy being microcanonically constrained and density being normalized.\cite{menon, patil-dunkel} Using Lagrange multipliers gives us the functional
\begin{equation}
    -\int_{ \Omega} \rho \log \rho \, d\mathbf{r} - \beta \lambda \int_{ \Omega} \psi \rho \, d\mathbf{r} - \alpha \int_{ \Omega} \rho \, d\mathbf{r}
\end{equation}
where $\psi$ is the streamfunction and $\rho$ is the distribution of vorticity. 
The functional has no direct dependence on any derivative of $\rho$, so to get an extremal we solve
\begin{equation}
    -1 - \log{ \rho} - \beta \lambda \psi - \alpha = 0
\end{equation}
We are working with $\lambda = 1$. Solving for $\rho$, then using the streamfunction-vorticity relation gives
\begin{equation}\label{one-species-mean-field}
    \nabla^{2} \psi = \lambda \rho = \frac{1}{ Z_{ \beta}} e^{\beta \psi},
\end{equation}
where $Z_{ \beta} = e^{1 + \alpha}$. We can use the normalization of $\rho$ to get an expression for $Z_{ \beta}$, which turns out to be a partition function\cite{pathria}:
\begin{eqnarray}
    \rho &=& \frac{1}{ Z_{ \beta}} e^{\beta \psi}, \nonumber \\
    \int_{ \Omega} \rho \, d\mathbf{r} &=& \frac{1}{ Z_{ \beta}} \int_{ \Omega} e^{\beta \psi} \, d\mathbf{r}, \nonumber \\
    Z_{ \beta} &=& \int_{ \Omega} e^{\beta \psi} \, d\mathbf{r}.
\end{eqnarray}

We can solve Eq.~(\ref{one-species-mean-field}) on the annulus if we assume $\psi$ exhibits angular symmetry. The equation becomes
\begin{equation}
    \psi''(r) + \frac{1}{r} \psi'(r) = \frac{1}{ Z_{ \beta}} e^{\beta \psi},
\end{equation}
with the boundary conditions $\psi(q) = \psi(1) = 0$. \\

The solution involves a series of variable changes\cite{menon}, the first of which is $t = \log{r}$, $\phi = \beta \psi + 2t$. Define $K = \frac{ \beta}{ Z_{ \beta}}$
\begin{equation}
    \phi'' = K e^{\phi}.
\end{equation}
Multiplying by $\phi'$ allows us to convert this into a first-order system, which can then be integrated to get:
\begin{equation}
    \pm dt = \frac{ d\phi}{ \sqrt{2K e^{\phi} - 2E}}.
\end{equation}
Where $E$ is taken to be a positive number. Making this first constant of integration negative is one of two choices that need to be made when solving the differential equation; why this is the correct choice for our current purposes is something that will be discussed shortly. \\

We next define $u = \sqrt{ \frac{E}{K}} e^{-\phi/2}$, which turns the above into
\begin{equation}
    \pm dt = \frac{ -2 u^{-1} \, du }{ \sqrt{ 2E u^{-2} - 2E } } = -\sqrt{ \frac{2}{E}} \cdot \frac{ du}{ \sqrt{1 - u^{2}}}.
\end{equation}
This then yields
\begin{equation}
    C \pm t = \sqrt{ \frac{2}{E}} \cdot \textrm{arcsin}{(u)}.
\end{equation}
Trace back the substitutions:
\begin{equation}
    u^{-2} = \frac{K}{E} e^{\phi} = \frac{ \beta}{E} \rho r^{2},
\end{equation}
which gives
\begin{equation}
    \rho = \frac{E}{ \beta} r^{-2} \sin^{-2}{ \left( \sqrt{ \frac{E}{2}} C \pm \sqrt{ \frac{E}{2}} \log{r} \right)}.
\end{equation}
Picking the plus sign, and renaming the parameters,
\begin{equation}
    \rho(r) = \frac{ 2E^{2}}{ \beta} r^{-2} \sin^{-2}{ \left( C + E \cdot \log{r} \right)}.
\end{equation}
We have two undetermined constants. One equation can be gained the normalization condition,
\begin{equation}
    1 = 2\pi \int_{q}^{1} r \rho(r) \, dr.
\end{equation}
The boundary conditions $\rho(q) = Z_{ \beta}^{-1}$, $\rho(1) = Z_{ \beta}^{-1}$ do not help us individually because we do not know what $Z_{ \beta}$ is, but we can combine them into the condition $\rho(q) = \rho(1)$. These conditions can be written as
\begin{equation}\label{mean-field-normalization}
    \beta = 4\pi E \cdot \big{[} \cot{(C + E \cdot \log{q})} - \cot{C} \big{]},
\end{equation}
and
\begin{equation}\label{mean-field-boundary}
    E = \frac{ -C + \arcsin{ \frac{ \sin{C}}{q}} }{ \log{q}}.
\end{equation}
This form is chosen to suggest a method of solution. Rather than start with $\beta$, one can pick $C$ arbitrarily, use Eq.~(\ref{mean-field-boundary}) to solve for $E$, and then use Eq.~(\ref{mean-field-normalization}) to solve for $\beta$. \\

We can now discuss the sign choices. The choice of the sign of $E$ changes the solution between having a sine or a hyperbolic sine. The sign choices for $t$ as well as the one that arises in solving the ``boundary condition'' for $E$ are the same choice, and this choice gives rise to different values of $\beta$ appearing as possibilities. The solution depicted in the derivation is the one that allows for large positive $\beta$. \\

\begin{figure}
    \centering
    \includegraphics[width=\columnwidth]{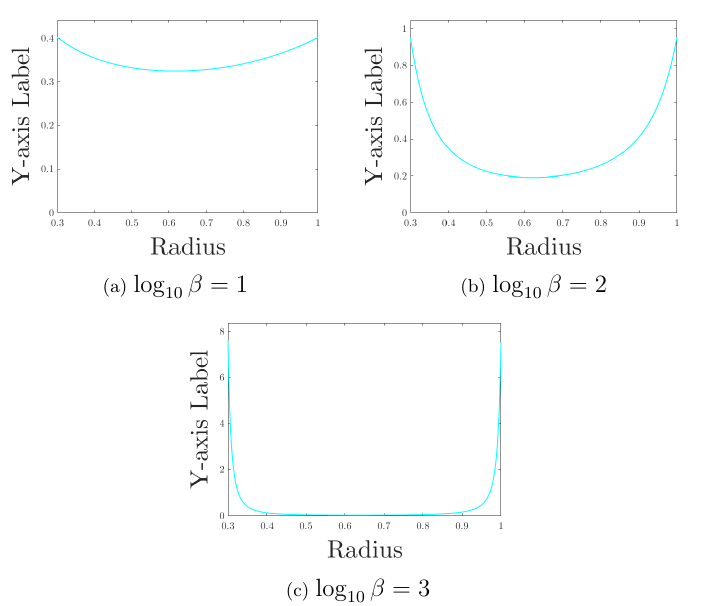}
    \caption{Mean field densities for selected $\beta$}
    \label{fig:mean-field}
\end{figure}

Some solutions of the mean field equation have their densities plotted in Fig.~\ref{fig:mean-field}.

\section{Empirical Microcanonical Results}

We can test the mean field distribution by sampling random distributions of vortices. The number of vortices for each sample was set at 40. To start we generated $5 \cdot 10^{5}$ random samples with each vortex being idependent and uniformly distributed on the domain; this allows one to map out the entire energy distribution. This distribution was used to determine which vortex configurations could be considered low temperature and thus simulated, with the threshold was set at the .05 quantile of energy. \\

A total of 300 configurations were generated and each one was simulated according to the Hamiltonian dynamics for 1000 time steps. The location of each vortex was recorded at each time step, and all of these locations were compiled into a density plot. \\

\begin{figure}
    \centering
    \includegraphics[width=\columnwidth]{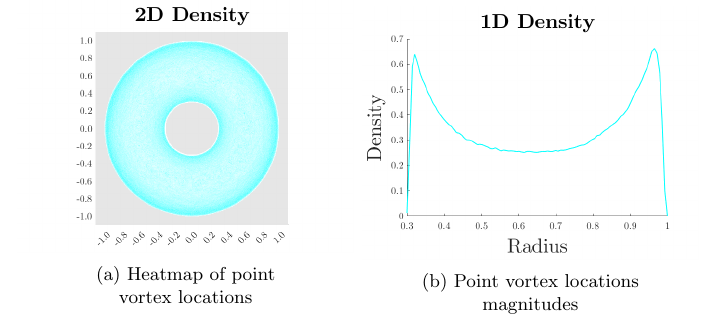}
    \caption{Microcanonical density}
    \label{fig:micro}
\end{figure}

As with the the mean field densities found analytically, the resulting distribution concentrates along the walls and is thinner in the interior. The thin layers of lower density near the domain walls in Fig.~\ref{fig:micro} are an artifact of the simulation, in which every time a vortex falls outside of the domain it is multiplied by a positive number such that it is set back inside the domain. The results otherwise agree with the behavior of the mean field solution, that density as a function of radius is convex. \\

\section{Lattice Model with Metropolis-Hastings}

A complementing numerical test that can be done with this system is to use a lattice model. The lattice model works with a fixed set of points, or equivalently with the Voronoi diagram formed from those points, and vorticity moving between them according to the Metropolis-Hastings algorithm, which allows us to sample from a canonical distribution defined by the parameter $\beta$ used by the algorithm.\cite{lim-nebus, atmospheres-book} \\

Metropolis-Hastings\cite{metropolis, hastings} works by proposing a perturbation to the system at each time step and calculating the difference in energy $\Delta E$ between this candidate state and the current state. A random number is sampled from $U(0,1)$ and compared to $\exp{( -\beta \Delta E)}$; if the random number is less than this value then the move is accepted; otherwise it is rejected. Notice that for positive $\beta$, if $\Delta E < 0$ then $\exp{( -\beta \Delta E)} > 1$ and the move is guaranteed to be accepted. However, because the energy distribution is unimodal, as energy decreases it becomes less likely that moves will be proposed which reduce it further still, and unless $\beta$ is very large some of the energy-increasing moves will be accepted. In this way, the algorithm's long-term behavior is for energy to fluctuate around but not converge to some average value which depends on $\beta$. The algorithm must be ran long enough that the system's initial state is washed out and the rolling average of energy is not moving monotonically. \\

A state of the system consists of values for each cell, where each value is the average vorticity across that cell; the overall system is normalized to have the integral of vorticity equal to $1$\cite{lim-nebus}. Because edge states are of interest, we originally experimented with grids not of uniform fineness, which included more points near the boundaries of the domain. The energy is computed as
\begin{equation}
    E = \frac{1}{2} \sum_{a,b = 1}^{N} M_{a,b} s_{a} s_{b},
\end{equation}
where $s_{a}$ and $s_{b}$ are the vorticities in a given site at a particular time step. The numbers $M_{a,b}$ are related to the average interaction between two cells. For $a \neq b$
\begin{equation}
    M_{a,b} = A_{a} A_{b} \cdot G(z_{a}, z_{b}),
\end{equation}
where $A_{a}$, $A_{b}$ are the areas of cells $a$ and $b$ respectively, and $z_{a}$, $z_{b}$ are lattice points. \\

The approximation of the average interaction strength between two cells by $G(z_{a}, z_{b})$ works well enough for two distinct cells, but the diagonal elements $M_{a,a}$ must be approached differently. For each cell, a fixed number $m = 120$ of points $z_{j}$ are sampled uniformly using a rejection method; we then computed
\begin{equation}
    \frac{1}{ m^{2}} \sum_{ j=1}^{m} \sum_{k=1}^{j-1} G(z_{j}, z_{k}),
\end{equation}
which approximates the average interaction between points in that cell. This value is related to the Robin function, but also depends on the cell size - as the lattice becomes finer the logarithmic singularity of the Green's function becomes more important and the system is high energy if vorticity is concentrated in a small number of cells, regardless of the value of the Robin function in those cells. \\

For each lattice, the cell areas were computed once, by forming a uniform square grid over the entire domain and counting how many points fell into each cell. The sampling used to populate the diagonal elements of $M$ was performed $50$ times. The results of this were paired with $50$ initial conditions, generated by sampling a uniform random number in each cell then normalizing the resulting distribution to have an integral of $1$. Each of these setups was ran on each value of $\beta$. \\

Notice that within each run, all of the necessary calls to the Green's function have been computed at the beginning, in contrast to the continuum approach where $O(N^{2})$ calls to the Green's function must be made at every time step. \\

An individual candidate move in our implementation consists of selecting two random cells and moving a uniform random amount of vorticity from one to the other. A single unit of time consists of as many candidate moves as there are points in the lattice. Each time the algorithm was ran, it was ran for 100 units of time. A burn-in period of 40 was uniformly applied, based on examination of energy plots - while the highest values of $\beta$ take approximately 40 units of time to reach their average energy, the middling and low range values of $\beta$ reach equilibrium in less than half this time. The principle result of the algorithm was taken to be the average distribution, with one sample taken at every unit of time. We can study how concentrated these distributions are by looking at entropy, the mean absolute deviation, and the variance. \\

\section{Lattice Results}

\begin{figure}
    \centering
    \includegraphics[width=\columnwidth]{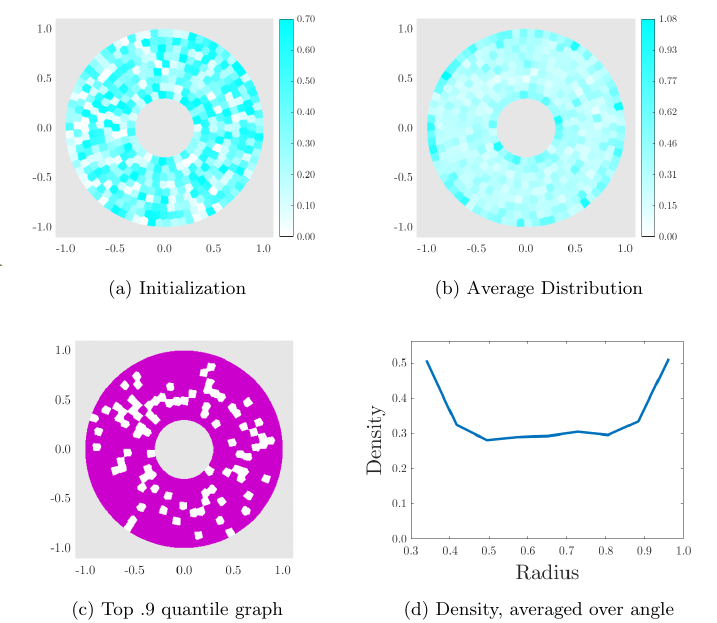}
    \caption{Data for lattice 5 ($m = 473$), initial condition 1, $\beta = 10^{0.75}$}
    \label{fig:5_075_1}
\end{figure}

\begin{figure}
    \centering
    \includegraphics[width=\columnwidth]{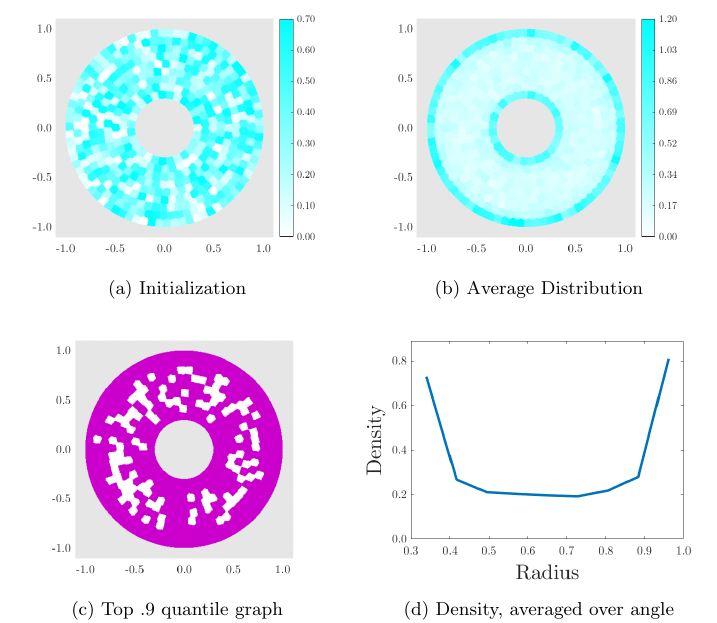}
    \caption{Data for lattice 5 ($m = 473$), initial condition 1, $\beta = 10^{2.00}$}
    \label{fig:5_200_1}
\end{figure}

There are 6 lattices, and 50 initial conditions for each one, generated through random sampling followed by normalization. Each lattice with each initial condition was ran with $\beta = 10^{k}$, with $k$ ranging in intervals of $\frac{1}{4}$ from $0$ to $5$. Lattices 1, 2, 3, and 4 were generated by a method that allows for smaller cells near the domain boundaries, although the parameters for lattice 2 were chosen so as to have close to equally sized cells. Lattices 5 and 6 were generated by a different method which has cells of approximately equal area all throughout the domain. The lattices have 442, 917, 2808, 3705, 473, 2334 cells respectively, although the number of cells does not appear to have much effect on the results. The initial condition and average distribution of a particular run are shown as an example in Fig.~\ref{fig:5_075_1}, as is a quantile plot generated by placing the cells in descending order of density, and filling in cells until 90\% of the average vorticity distribution has been accounted for. The same data is shown for the same lattice and initial condition, but a lower temperature, in Fig.~\ref{fig:5_200_1}. \\

\begin{figure}
    \centering
    \includegraphics[scale=0.5]{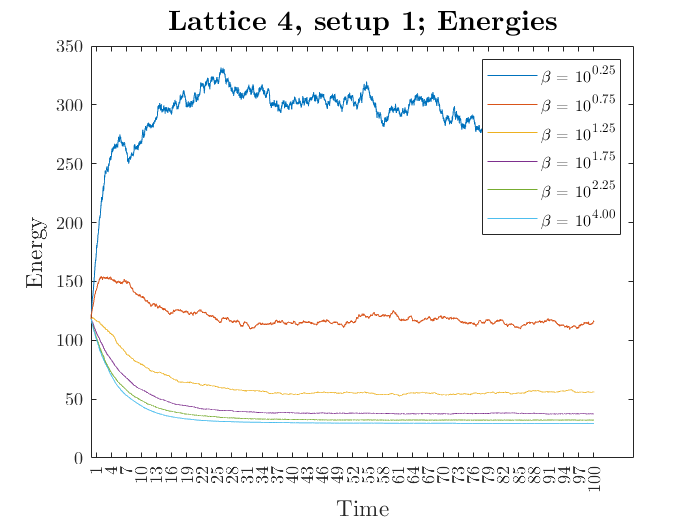}
    \caption{Lattice 4 ($m = 3705$), initial condition 1, energies for $\beta = 10^{0.25}, 10^{0.75}, 10^{1.25}, 10^{1.75}, 10^{2.25}, 10^{4.00}$}
    \label{fig:lattice-4-energy}
\end{figure}

\begin{figure}
    \centering
    \includegraphics[scale=0.5]{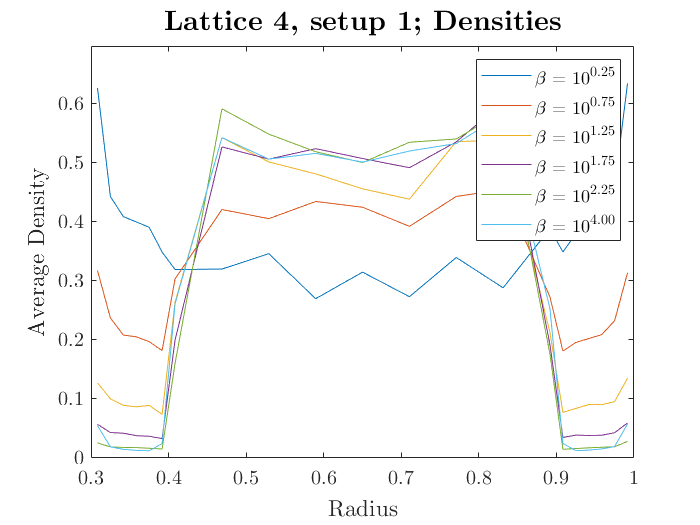}
    \caption{Lattice 4 ($m = 3705$), initial condition 1, radial densities for $\beta = 10^{0.25}, 10^{0.75}, 10^{1.25}, 10^{1.75}, 10^{2.25}, 10^{4.00}$}
    \label{fig:lattice-4-density}
\end{figure}

\begin{figure}
    \centering
    \includegraphics[scale=0.5]{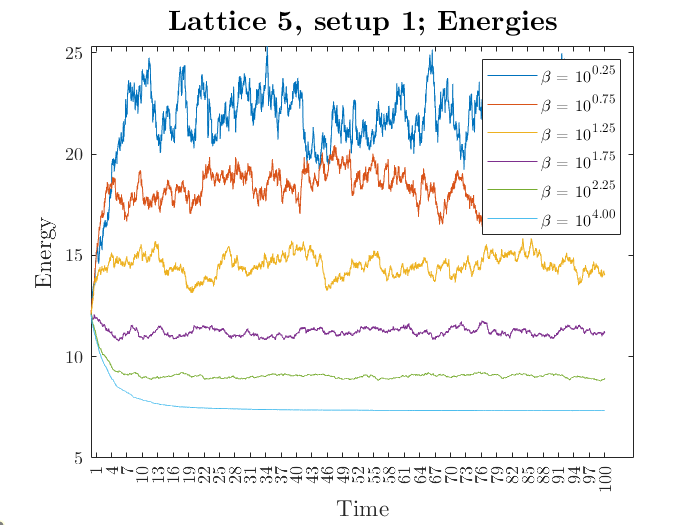}
    \caption{Lattice 5 ($m = 473$), initial condition 1, energies for $\beta = 10^{0.25}, 10^{0.75}, 10^{1.25}, 10^{1.75}, 10^{2.25}, 10^{4.00}$}
    \label{fig:lattice-5-energy}
\end{figure}

\begin{figure}
    \centering
    \includegraphics[scale=0.5]{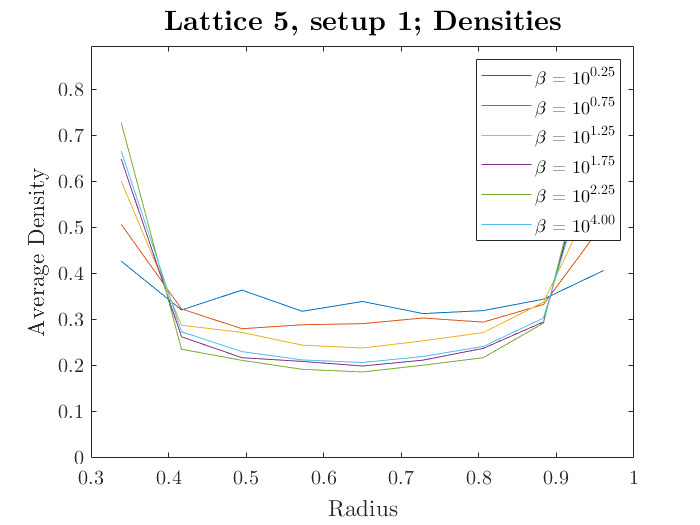}
    \caption{Lattice 5 ($m = 473$), initial condition 1, radial densities for $\beta = 10^{0.25}, 10^{0.75}, 10^{1.25}, 10^{1.75}, 10^{2.25}, 10^{4.00}$}
    \label{fig:lattice-5-density}
\end{figure}

Energy over time and radial density for certain values of $\beta$ for lattice 4, configuration 1 are shown in Fig.~(\ref{fig:lattice-4-energy}) and Fig.~(\ref{fig:lattice-4-density}). Corresponding values for lattice 5, configuration 1 are shown in Fig.~(\ref{fig:lattice-5-energy}) and Fig.~(\ref{fig:lattice-5-density}). \\



\begin{figure}
    \centering
    \includegraphics[scale=0.5]{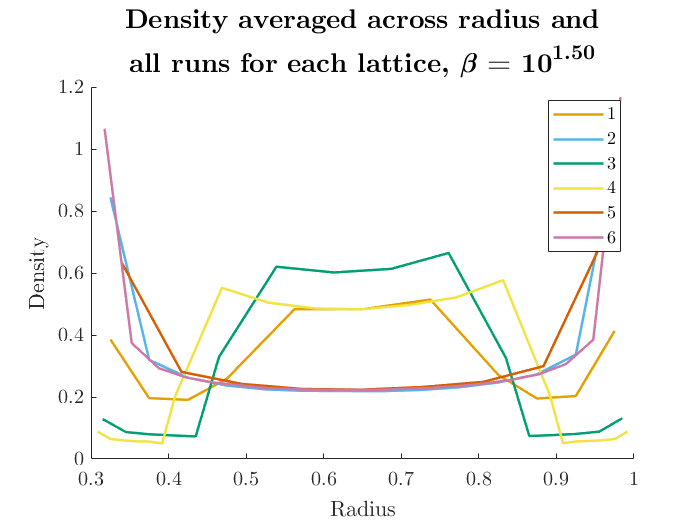}
    \caption{Radial density for $\log_{10}( \beta) = 1.50$, averaged across all 50 runs, for each lattice}
    \label{fig:overall}
\end{figure}

Large positive $\beta$ always leads to low energies, and in the continuum model low energy states of the system are associated with vorticity being concentrated near the edge. We continue to see this behavior in the lattice model much of the time. However, the lattice model also exhibits certain discretization artifacts. The lattice places a limit on how close to the boundary vorticity can be considered as being. This is in stark contrast with the continuum model, in which energy reductions can always be made by moving vortices some infinitesimal amount closer to the boundary. Moreover, the assumption that vorticity is uniformly distributed within each cell means that there are effects mediated by cell size, in which larger cells have lower self-interaction than smaller cells. For the lattices we generated that had larger cells in the interior of the domain than near the walls, this led to degenerate behavior in which vorticity preferred to be in the interior cells, with a small uptick in concentration at the boundary, visible in e.g. lattice 4, see Fig.~(\ref{fig:lattice-4-density}). This behavior is also seen in lattices 1 and 3, but not in 2, 5, or 6 with their approximately equi-sized cells. See Fig.~(\ref{fig:overall}). For these lattices there is still a preference for vorticity to be located in larger cells, but the effect is much weaker. This was tested by picking out Voronoi cells whose underlying points had the same magnitude and calculating the Spearman coefficient for cell size against average concentration. The results were small but consistently positive ($0-0.25$), confirming the relationship. \\

\begin{figure}
    \centering
    \includegraphics[scale=0.5]{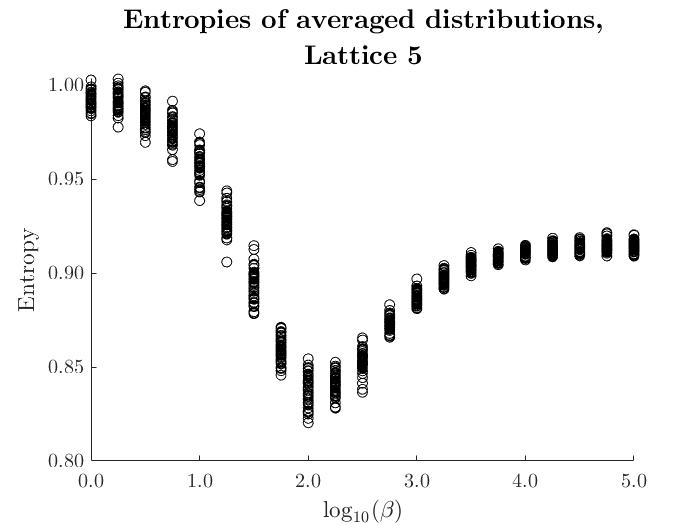}
    \caption{Entropy of average distribution for all runs (50 for each $\beta$) on lattice 5 ($m = 473$)}
    \label{fig:entropies-5}
\end{figure}

\begin{figure}
    \centering
    \includegraphics[scale=0.5]{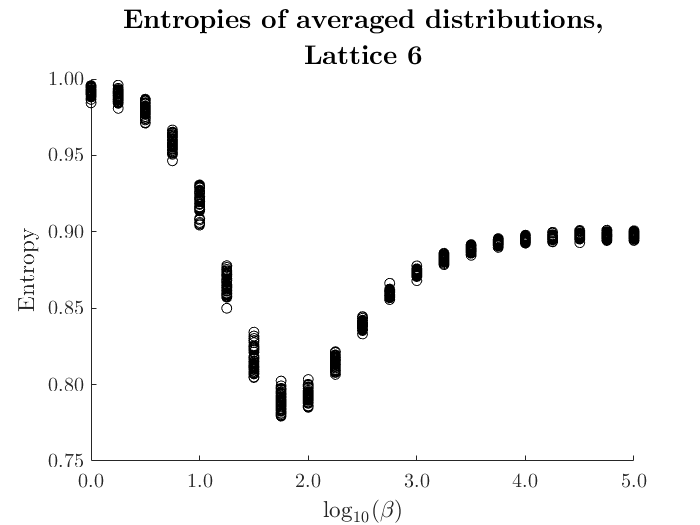}
    \caption{Entropy of average distribution for all runs (50 for each $\beta$) on lattice 6 ($m = 2334$)}
    \label{fig:entropies-6}
\end{figure}

The concentration of the distribution, as measured by entropy, variance, and mean absolute deviation, goes up with $\beta$ for $\beta$ in the range $10^{0} - 10^{2}$. We can also this graphically by comparing e.g. the quantile graphs in Fig.~\ref{fig:5_075_1} (c) and Fig.~\ref{fig:5_200_1} (c). After this the distribution becomes less concentrated for increasing $\beta$. This can be seen in Fig.~(\ref{fig:entropies-5}) and Fig.~(\ref{fig:entropies-6}). The model itself features a negative correlation between energy and entropy, which is believed to be the cause of this behavior.


\section{Conclusion}

These findings show that low energy edge states for the vortex gas can be found in a multiply-connected domain. It was also found that in order to run lattice models with large $\beta$, one must be mindful of effects mediated by the areas of the cells involved. \\

The most direct implications of these experiments involve turbulence in bounded plasma systems such as fusion reactors and plasma etching devices. Future works may wish to examine domain geometries which are multiply connected but unbounded, which would have relevance to atmospheric or astrophysical phenomena, or which have multiple vortex species present. Furthermore one might examine how the connection between vortex models and Chern numbers found by \cite{patil-dunkel} changes with alternate vortex species or domains.


\section*{Conflict of Interest Statement}
The authors have no conflict of interest to report.

\section*{Data Availability Statement}
Data available upon reasonable request

\nocite{*}
\bibliography{sources}

\end{document}